\documentclass[lineno,10pt]{aseecoed}

\articletype{Original}

\title{Enhancing Computer Science Education with Pair Programming and Problem Solving Studios}

\author[1][jorr@georgefox.edu]{J. Walker Orr}
\affil[1]{Electrical Engineering and Computer Science, George Fox University, Newberg, OR, 97132, USA}

\publicationdate{TBD}
\volume{TBD}
\issue{TBD}
\submitteddate{5 June 2023}
\accepteddate{28 Sept 2023}
\citethisas{Orr (2023) Enhancing Computer Science Education with Pair Programming and Problem Solving Studios. aseecoed.}{}

\begin{document}

\nolinenumbers

\maketitle

\begin{abstract}
This study examines the adaptation of the problem-solving studio to computer science education by combining it with pair programming.
Pair programming is a software engineering practice in industry, but has seen mixed results in the classroom.
Recent research suggests that pair programming has promise and potential to be an effective pedagogical tool, however what constitutes good instructional design and implementation for pair programming in the classroom is not clear.
We developed a framework for instructional design for pair programming by adapting the problem-solving studio (PSS), a pedagogy originally from biomedical engineering.
PSS involves teams of students solving open-ended problems with real-time feedback given by the instructor.
Notably, PSS uses problems of adjustable difficulty to keep students of all levels engaged and functioning within the zone of proximal development.
The course structure has three stages, first starting with demonstration, followed by a PSS session, then finishing with a debrief.
We studied the combination of PSS and pair programming in a CS1 class over three years.
Surveys of the students report a high level of engagement, learning, and motivation.
\end{abstract}



\section{Introduction}

Pair programming is an eXtreme programming (XP) methodology \citep{beck2000extreme} that has seen some use in industry \citep{hannay2009effectiveness}.
It involves two programmers working together on a single problem and computer with one programmer taking the role of a ``driver'' and the other in the role of ``navigator.''
The ``driver'' operates the keyboard and directly writes the code while the ``navigator'' observes and asks questions, critiquing and refining the code and its design.
The ``navigator'' is not passive, they watch for bugs and defects, think of alternative designs, and look up related documentation and resources.
Though the effectiveness of pair programming is mixed \citep{hawlitschek2022empirical, hannay2009effectiveness}, in some cases it has been shown to produce higher quality code faster than solo programming \citep{williams2000strengthening}.
The intention is that pair programming will help developers working together catch mistakes and defects much faster than on their own.

For education, pair programming is compelling because it fits into the paradigm of apprenticeship and distributed learning, the idea that ``Knowledge is commonly socially constructed, through collaborative efforts toward shared objectives or by dialogues and challenges brought about by differences in persons' perspectives'' \citep{salomon1997distributed}.
Further, it has been shown to increase student satisfaction, reduce student frustration, improve student's tendency to persist, and give students a sense of self-efficacy \citep{williams2001support}. 
The ICAP framework describes four modes of student engagement and behavior, identifying the interactive mode as producing the highest level of student cognitive engagement.
Interactive modes of learning are believed to produce deep, transferable knowledge \citep{chi2014icap}.
Pair programming fits within ICAP's definition of interactive learning and hence has the potential to produce robust, transferable, conceptual learning.
Recently, \cite{hawlitschek2022empirical} conducted a literature review and meta-study of pair programming in education concluded that pair programming is important and effective for students, especially beginners, but effective instructional design was missing.
Hence pair programming has been shown to have a lot of potential as a teaching methodology but the details of how to implement it correctly in a classroom has yet to be discovered.

We propose that the solution to effective instructional design for pair programming in the classroom has been found in the Problem Solving Studio (PSS) learning environment \citep{pss}.
PSS was designed to teach biomedical engineering students to solve complex problems without having to resort to rote memorization of procedures and algorithms.
Students work in teams of two to solve ill-defined problems in a public space, enabling instructors to provide real-time feedback as they progress.
A key feature of PSS is \emph{dynamic scaffolding}, a targeted adjustment of problem difficulty to keep students challenged but not discouraged.
By increasing or decreasing the difficulty on a per-team basis in real-time, as many students as possible can be kept in the zone of proximal development.
A lecture-based course will have a difficult time matching the variety of levels that students are at since the same lecture content and delivery are communicated to all the students.
There is good evidence that PSS improves students' conceptual understanding \citep{pss}.

PSS and pair programming are a natural fit and the combination of the two match the objectives and pedagogical needs of CS1 courses.
For this reason, this study specifically addresses the adaptation of PSS in conjunction with pair programming for CS1 pedagogy.
Two key objectives of a CS1 course are to teach algorithmic problem solving skills and a specific programming language.
One of the challenges for expert instructors is that both problem solving and language knowledge is so deeply ingrained that it is second-nature to the instructors.
Paradoxically, this high level of understanding means instructors often have a difficult time communicating this knowledge since it is taken for granted \citep{pss}.
Further, student ability and background varies significantly in CS1 courses.
However, the synthesis of PSS and pair programming addresses these challenges and objectives directly, by giving students a hands-on opportunity to develop problem solving and programming language skills.
PSS with the addition of pair programming, teaches algorithmic problem solving through a cognitive apprenticeship environment \citep{collins1987cognitive}.
Students learn from each other and are also guided by the instructor or other teaching assistants.
Peer learning is helpful since students who are at similar levels of ability have recent experience with similar problems.
This means that they are often better at communicating those solutions since they remember the details and particularities of both what they found challenging and how they overcame those obstacles. 
Pair programming's ability to give rapid feedback helps students learn the syntax and semantics of a programming language.
Further it promotes pair collaboration and problem solving.
Pair programming has been shown to help notice programmers solve problems can not handle on their own \citep{hawlitschek2022empirical}.
The combination of PSS and pair programming creates both an apprenticeship and peer learning environment in which students develop both problem solving and programming language skills.





\section{Literature Review}

There has been a significant amount of research into using pair programming for educational purposes.
The popularization of pair programming largely started by \textit{Extreme Programming Explained} \citep{beck2000extreme}.
It promoted pair programming, among other techniques, as a way of producing high quality code faster than conventional methods.
\cite{williams2001support} combined professionals with advanced undergraduates and found that ``Experimental results show pair-programming pairs develop better quality code faster with only a minimal increase in pre-release programmer hours.''
Further, they noted that the programmers generally found that pair programming was more enjoyable than programming alone.

\cite{hannay2009effectiveness} studied the effect of using pair programming in a CS1 course with 600 students.
Students did their work with pair programming, alternating driver and navigator roles.
The results were better quality programs and a significantly higher rate of completion for the course among those that participated in pair programming.

\cite{williams2001support} had students use pair programming for their work in a web based programming course.
The results of an anonymous survey conclude that the 74\% of students thought that they could solve any problem with their partner and 84\% believed they learned faster because of their partner.

Though there are some mixed results among studies as well.
\cite{mcdowell2003experimenting} saw no improvement in midterm or programming scores, but did see an improvement under some ``holistic'' scores.
A large randomized trial of 1,530 undergraduates found that pair programming was virtually unrelated to all measured outcomes.
However, pair programming did appear to hurt the grades, success, and likelihood of White students and had no effect on any other demographic \citep{bowman2020pair}.
A study of persistence of CS students suggests that pair programming might improve the persistence of women in CS though the results were not statistically significant \cite{werner2004pair}.

\cite{hannay2009effectiveness} conducted a meta-analysis of 18 studies concluded that has an overall positive impact on software quality and time, but did require more total effort by the programmers involved.
The studies were on a mix of educational and professional teams.
However, they noted that pair programming offers a substantial benefit for junior developers, with their code correctness increasing 73\% on regular programs and 149\% on more complex programs.

Recently another meta-study was conducted on all the published articles on pair programming for higher education from 2010 to 2020.
The analysis contained 61 articles and contained some important conclusions and observations.
First, pair programming is generally beneficial for student learning, particularly for inexperienced students.
Second, pair programming is generally difficult to apply successfully in the classroom.
They noted a lack of research on effective instructional design for pair programming.
Most research on instructional design is on how to select the pairs, though the results are mixed.
In particular, \cite{hawlitschek2022empirical} identified that the problem with pair programming is when two weak students are paired together without much guidance.

\section{Approach}
PSS is a ``flipped classroom'' pedagogy in which pairs of students work together to solve ill-structured, complex problems all while engaging in a critical dialog with the instructor.
This problem solving centered approach is intended to match real-world problem solving and encourage deep, conceptual thinking, and partner-based learning.
As the pairs work on a problem, they will receive on the spot feedback called a desk crit.
In addition, the instructor may feel the need to adjust the problem difficulty up or down to match the student's ability and progress \citep{pss}.
This real-time adjustment is called a ``dynamic scaffold'' and is used to keep the students in their zone of proximal development (ZPD) \citep{vygotsky1978mind}.
Overall, PSS is a cognitive apprenticeship model of instruction, in which students learn by practicing problem solving and receiving personalized feedback as they work.

Our method is to adapt PSS to use the practice of pair programming.
In our adaption, a series of shorter problems was found to be the most effective approach.
The series of problems keep the pairs on-track, focused, and within their ZPD.
This also provides more opportunities to employ the dynamic scaffolding to adjust problem difficulty.


\subsection{Problem Structure \& Dynamic Scaffolding}

One essential aspect to PSS is the problem structure and formulation.
Rather than using typical textbook problems, PSS challenges students by presenting them with problems that are more ill-structured and complex \citep{pss} according to \cite{jonassen2015all}'s problem difficulty scheme .
These problems are designed and intended to have multiple legitimate solutions \citep{jonassen2015all}.

In the original PSS for engineering, a single class was typically two hours in length.
During a single class, a student team was expected to finish between one and three problems \citep{pss}.
This illustrates how the problems are lengthy and challenging with possibly many ``deadends'', sub-problems, and an overall meandering path to a solution.
By comparison, typical textbook problems are far more formulaic, straightforward, and linear.

In our adaptation of PSS for CS education, the problems are typically presented to the students with examples of program inputs and desired outputs.
For example, if the problem is to write a program to produce prime numbers, the prompt for the student is minimal ``Work with your partner to write a program that produces prime numbers.''
The prompt is paired with example input and output, in this case a command-line interface:

\begin{verbatim}
Enter the upper limit for primes: 10
The primes under 10:
2
3
5
7
\end{verbatim}

This example is sufficiently ill-structured and complex for a CS1 course because it is solvable with their knowledge and skill level and allows for multiple legitimate solutions.
First, the pair must identify the problem and challenges that need to be overcome to produce a solution.
The decisions about which control structures to use and how to use them or which functions would be helpful to define or use are entirely up to the student pair.
Each pair must create a complete solution i.e. a program for the given problem from scratch.
Further, students may have to do some independent research on the particular application area, in this example, on the relevant properties of prime numbers to complete the task.
In general, an example of input to a program and corresponding desired output is essentially a bare minimum specification for a program.
In that sense, the problems are presented in an ill-structured fashion, since they do not explicitly ask for particular programming constructs or methodologies, only a desired goal.
This gives the student pairs the opportunity to navigate a large space of solutions while keeping all the students in the class on the same task.

Another key aspect of PSS problems is complexity.
The problem at the center of a PSS is targeted at the more advanced students.
This ensures a sufficiently complex problem, however it will be too challenging for many of the students.
The way this issue is addressed is through ``dynamic scaffolding'', the process of instructors adjusting problems in real time \citep{pss}.

If a team is stuck and unable to make any progress on a problem, the instructor may adjust the problem difficulty down by making the problem less complex, more structured, or both.
This allows as many students as possible to be within their ZPD \citep{vygotsky1978mind}.
The ZPD means a challenge is appropriate for student learning, meaning that it is neither too challenging to be discouraging nor too boring to be uninteresting.
Dynamic scaffolding means students are far more likely to be in their ZPD than traditional lectures which are typically only suited for a subset of students.

Our application of PSS for CS extends this idea of dynamic scaffolding to the problem selection and formulation.
Rather than selecting a single problem for the PSS, our adaptation utilizes between one and five problems.
The typical structure of our CS PSS is to have a ``ladder'' of problems, starting with a relatively easy problem, then followed by moderately difficult problems, and then finally ending with an advanced problem.
All the problems are centered around a single concept or methodology being taught, for example, looping control structures.

We believe this ``problem ladder'' has some advantages over a single advanced problem.
First it allows students to work at their own pace and naturally fall into their ZPD.
Students will typically have both the experiences of successfully solving a problem and the experience of being challenged by another.
This way students are encouraged by solving a problem and are also made aware that there is still more to learn.
This is particularly important in CS1 classes since students typically have a diverse background regarding prior programming experience.
It is important to encourage and motivate the inexperienced students while challenging the move advanced students.
One important aspect of the ``problem ladder'' is that instructors should clearly communicate that students are not expected to solve all the problems.
This will help prevent under-performing students from becoming discouraged by advanced problems.

Each problem on the ``ladder'' can be dynamically adjusted as well, which enables the instructor to have a fine-grained control over problem difficulty.
For example, if a pair spent most of the class period solving the ``easy'' problem, it is often more useful for them to revise an adjusted version of the problem in the remaining time.
That way they can spend their time focused on solving the problem rather than on the contextual switch to another problem.
This dynamic adjustment is useful at both ends of the ``ladder''.
If the ``easy'' problem is too difficult, it can be adjusted down.
Likewise, if a pair of advanced students solve all the problems in the ladder, the last problem can be adjusted to be more difficult.
Returning to the example of computing prime numbers, one way to apply dynamic scaffolding to the problem is to change the requirement of producing all the primes under a limit to instead write a function that determines if a given number is prime or not.
The complexity of the problem is reduced because the initial version of the problem requires two loops to solve, while the revised version only requires one.
Also this provides additional structure to the problem since the goal is made more specific.

To ensure student motivation, participation in the PSS is part of the course grade.
Though the requirement is not onerous, all that is required of the students is their prescience and a good-faith effort.
Attendance to PSS sessions are obviously key to student learning and the requirement to put in some effort helps encourage them to climb up the problem ladder.
Though, the natural progression of problem difficulty and early success provides intrinsic motivation.
In practice we observed that students enjoyed solving problems and were naturally motivated generally speaking.

\subsection{Informal Assessment \& Feedback}

PSS applies the idea of a desk crit from architecture design studios, where instructors give ``informal formative assessment'' to students through a discussion of their work \citep{dinham1987ongoing}.
In the PSS framework, instructors provide this feedback and assessment through asking unobtrusive, open-end questions \citep{pss}.
For example, the dialog can be initiated by asking questions such as ``How are you doing?'', ``What are you working on now?'', or ``Are you making progress?'' can open a dialog that can provide specific instruction, assistance, or feedback to the pair.
The initial question enables the instructor to quickly determine the status of the student pair.
From there more intentional and specific questions can be asked to promote deeper thinking \citep{raths1967teaching}.
These dialogues are effective opportunities to discuss how the students solved a problem, what issues they are stuck on, alternative solutions, get help on practical issues, starting problem solving, conventions regarding design and communication of solutions, ways to an improve a solution, clarify conceptual misunderstandings, and so on.
Also, it is natural during these dialogues to enact dynamic scaffolding by adjusting the problem difficultly up or down based on how the pair is performing.
Further, these dialogues are a good time for individualized help.
Naturally over both the class period and the course of the semester, the need for support diminishes and the pairs work effectively on their own.
Instructors can also get a sense of which topics or methodologies the entire class is struggling with versus individual students.
This feedback for the instructor can help them improve their lectures or demonstration sessions as well as the PSS sessions themselves.
This means the course design can be adjusted on-the-fly or improved for the future.
If many pairs are struggling with the same issues, the instructor can stop the PSS and give some brief instruction or clarification to the entire class.
What this feedback provides is a means for the instructor to identify ``troublesome knowledge'' that the students are struggling to learn.
With these insights, the instructor can provide targeted instruction and clarification to specifically address these difficult topics \citep{perkins2006constructivism}.

In our adaptation of PSS for CS, we incorporated the desk crit in two different ways.
First, these dialogues are engaged informally as the student pairs work on problems.
More time is focused on the pairs that were struggling the most to provide extra support.
For the more advanced students, desk crits are more of a time to suggest improvements or alternative solutions. 
Secondly, the desk crits are employed when a pair finishes a problem.
This is a good time to provide extra encouragement, feedback, improvements, etc.
Further, it is a good way to informally assess student progress and ability by tracking both the number of problems a pair finishes and the amount of time they need.

In general, the desk crit is a good way to provide individualized instruction.
Struggling students can get help with the practical or conceptual problems they have.
More advanced students can be challenged to think more deeply about the problem or improve the presentation or design of their solution.
This kind of individualized instruction corresponds well to the student's ZPD.

Naturally an instructor can only provide feedback to one team at a time.
To handle larger class sizes, teaching assistants can be utilized to give feedback and other guidance to student pairs.
This was notable part of the original PSS design \citep{pss}, however did not find the use of teaching assistants necessary for our adaptation.

\subsection{Pair Programming Dyad}

An essential component of PSS is the grouping of students into pairs called dyads \citep{pss}.
The dyads are important since it gives each student a partner to better navigate, interpret, and solve the problem \citep{hutchins2020distributed}.
This partnership facilitates students sharing knowledge and expertise as well as providing each other with helpful critique.
Moreover, the instructor's knowledge and skills, having reached an expert level of ability, are tacit and difficult to articulate.
Students however are at closer levels of ability to each other and are in the process of learning the material which can make them more suited to answering each other's questions.

In the original PSS, students shared a 17'' x 22'' blotter pad as a publicly visible problem solving space.
One student writes on the pad while the other partner observes, listens carefully, agrees or critiques the writer.
After a few minutes the pair exchanges the roles of writer and observer.
The students also negotiate the duration of the writer role \citep{pss}.

Similarly, in computer science education and in industry, pair programming is a well-established practice of team-based problem solving \citep{beck2000extreme, hannay2009effectiveness, hawlitschek2022empirical}.
The complementary pair of roles has been shown to be an effective way of producing high-quality programs and is generally enjoyable for both partners \citep{hannay2009effectiveness, williams2000strengthening}.
The adaptation of PSS to CS replaces the blotter pad and roles of writer and observer with pair programming, that is, a single computer and the roles of driver and navigator.
This closely matches this aspect of the design of the original PSS while contextualizing it to CS education.

A recent large meta-study concluded that the effectiveness of pair programming in education was mixed, but that was primarily due to a lack of effective instructional structure and guidance \citep{hawlitschek2022empirical}.
Our adaptation of PSS seeks to address this problem, specifically with the structure provided by the use of ill-structured, complex problems, dynamic scaffolding, problem ladders, and desk crits.

Solving ill-structured, complex problems is arguably the centerpiece of PSS.
For the driver, the benefits of working on these problems is simply the process of solving and writing a solution to the problem.
However, the navigator is particularly important and beneficial. 
Since the problems are ill-structured, by definition there is a lot of research that needs to be done.
As the driver writes the code, the navigator can research possible important components of a solution, key knowledge about the particular problem, refer back to the course textbook, consult examples, and explore alternative solutions.
Additionally, the navigator can observe the problem solving process of the driver. 
This is especially if the navigator is relatively inexperienced compared to the driver.
For most people, the process of problem solving is difficult to communicate.
The navigator has the opportunity to learn by example, which directly ties into the apprenticeship mode of learning that PSS encourages.
This means the navigator is learning from observing their partner, by listening to the feedback from the desk crit, and by participating and communicating with their partner.

The problem ladder gives direction for each pair and naturally provides opportunities for the partners to swap roles.
Our instructors communicated and emphasized that pairs exchange roles when they complete a problem.
This allows each member to get the experience and benefits of being both a ``driver'' and a ``navigator''.
Further, the exchange of roles helps prevent either partner from either dominating or withdrawing. 
Individuals who tend to dominate and be assertive will benefit by performing the thoughtful, observation-centered role of the navigator.
Likewise, individuals who tend to be more passive will benefit by performing the direct, active role of the driver.

One of the criticisms of pair programming for novice programmers is that it is ``the blind leading the blind'' \citep{rosenberg2008extreme}.
However, the combination of the problem ladder, desk crits, and dynamic scaffolding, student pairs that are struggling can be quickly identified and put back on a track to success.
Within a few minutes of starting the PSS activity, it becomes clear which pairs will be successful on their own and which need some help or guidance.
Further, if a large number of pairs are stuck, a quick demonstration or example by the instructor can get the class back on track.
The identification of struggling students is much quicker than a traditional lecture-based course.
Typically the only time when struggling students are identified is when they turn in their assignments which means it could take days or weeks.
In a PSS, these students can be found and helped within a single class period.



Both PSS and pair programming are centered on learning through the exchange of knowledge between peers in the context of a shared problem.
The educational benefit of pair programming is primarily from the verbalization of problem solving it encourages.
The details and roles of pair programming are not necessarily where the benefits come from but rather the interaction and discussion they facilitate \citep{hannay2009effectiveness}. 
This is likely the case for PSS as well, which is based on constructive learning \citep{ruiz2011informal}, the benefit of PSS is the facilitated dialog between students and between student and instructor.
This is the central idea of combining PSS and pair programming, to create dialog, both between students and between student and instructor, around problem solving for CS.
This dialog-centered mode of learning allows students to construct their own knowledge of CS through problem solving.





\begin{table}[]
    \centering
    \begin{tabular}{l c c c c c }
         \textbf{Year} & \textbf{Responses} & \textbf{PSS Useful} & \textbf{PSS Engaging} & \textbf{PSS Challenging} & \textbf{Partner Useful} \\
         \hline\\
         2020 & 34 & 3.97 & 100\% &  97.06\% & 94.12\%\\
         2021 & 43 & 4.23 & 97.73\% & 86.36\% & 97.73\%\\
         2022 & 39 & 4.18 & 100\% & 92.50\% & 92.50\%\\
         Combined & 116 & 4.16 & 99.15\% & 91.53\% & 94.92\%\\
    \end{tabular}
    \caption{Survey data for CS1 over three years. ``PSS Useful'' is on a 1 to 5 Likert scale with 1 indicating not useful and 5 indicating very useful. The other columns have a binary response and the value reflects the proportion of with a positive answer.
    The four corresponding survey questions are in order ``In general, are the in-class activities useful to you?'', ``Are the activities engaging?'', ``Are the activities challenging?'', and ``Did you appreciate having a partner?''.
    }
    \label{tab:201}
\end{table}

\subsection{Course Structure}


The course we implemented our application of PSS for CS in CS1, CS2, and web programming courses, but this study is focused on the application to CS1.
These courses followed the typical 3 credit hour format, 50 minute meetings 3 times per week.
A similar format was used for each course.
The first meeting for the week is a demonstration session.
A new concept or topic is discussed and applied by the instructor.
This typically means that the instructor solves a problem by writing a program and testing it.
For example, for a CS1 course, a topic could be for-loops.
The second meeting of the week is a PSS session.
Students are presented a ``ladder'' of problems related to that week's topic.
At a minimum, they are expected to solve the first rung on the `ladder' since all the problems are related to the topic.
The third meeting is either a debrief of the recent PSS, another PSS, or both.
The debrief consists of the instructor solving the ``second rung'' problem while explaining each step.
This gives students who only solved the first problem an example of how to extend their knowledge and skill.
For more advanced students, they have the opportunity to see an expert solve a problem in a way which is likely different from how they solved it.
From week to week (and even from course to course), some problems are revisited but with new concepts and skills available.
This gives an opportunity to see how the process and solution changes with the new material.
In summary, the weekly structure of the course is demonstration, PSS, and finally debrief.

PSS is a replacement for the traditional lecture-based pedagogy.
Other aspects of the course such as assignments and exams are independent of the PSS.
For our PSS adaptation, the same homework and exams were utilized.
In the CS1 course, there are 10 homework assignments, an online textbooks with built-in reading and programming assignments, 2 midterms, and a final exam.
This is the same structure of assignments and exams that was employed before the adoption of PSS.
However, the problems presented in the PSS parallel the student's homework assignments.
The students' experience with a type of problem proceeds as: demonstration, PSS, debrief, and finally homework assignment.
The homework is done individually, however the students have the opportunity to learn from their instructor and peers before engaging with the problem on their own.

\section{Discussion}

\begin{table}[]
    \centering
    \begin{tabular}{l c c c c c}
         \textbf{Year} & \textbf{Responses} & \textbf{Prefer PSS} & \textbf{Prefer Lecture} & \textbf{Prefer Both} & \textbf{Prefer Neither}\\
         \hline\\
         2020 & 34 & 35.29\% & 5.88\% & 58.82\% & 0.00\%\\
         2021 & 43 & 38.64\% & 4.55\% & 54.55\% & 2.27\%\\
         2022 & 39 & 40.00\% & 7.50\% & 52.50\% & 0.00\%\\
         Combined & 116 & 38.14\% & 5.93\% & 55.08\% & 0.85\%\\
          
    \end{tabular}
    \caption{Survey data on which mode of instruction students prefer. The question presented was ``Which do you prefer?'' The options for response were ``Lectures,'' ``In-class exercises and activities,'' ``Both,'' and ``Neither.'' }
    \label{tab:201_pref}
\end{table}

Our adaptation of PSS for CS was implemented in a CS1 course and evaluated over a three year period.
In order to evaluate our implementation, voluntary anonymous surveys were conducted to assess the student's perception of its usefulness.
Students had two opportunities to fill out the survey over the course of the 15 semester at about weeks 5 \& 10.
The results on how students viewed the PSS sessions can be found in Table \ref{tab:201}.
Overall the results of the surveys are very positive.
Averaged over the three years, 99.15\% found PSS to be engaging, 91.53\% found it to be challenging, and 94.92\% thought their partners were useful.
Furthermore, the average score of 4.16 for PSS usefulness, on a Likert scale of 1 to 5, strongly supports the claim that the PSS was helpful for student learning.
This is notable considering the diverse backgrounds of the students in terms of exposure to CS and computer programming.
At our institution, CS1 is a formal requirement or strongly encouraged by a variety of majors including accounting, biology, finance, and engineering in addition to computer science.
These results appear to strongly suggest that the combination of the problem ladder and dynamic scaffolding generally kept students challenged and in their ZPD.
Given the overall positive responses, the PSS must have been effective at matching the appropriate level of difficulty for the majority of students.
It is worth noting that since the data was collected via voluntary, anonymous surveys, that this is a caveat to their strong results.
Since the students had two opportunities to fill out the survey and the responses are anonymous, the statistics reported in Tables \ref{tab:201} \& \ref{tab:201_pref} likely aggregate over multiple responses from a single student.
Also, based on the number of responses, there were some students who completed the courses but did not fill out the survey.
However, the survey results do match the anecdotes from instructors that students are generally very engaged, asking good questions, completing problems, and are apparently enjoying the experience.


Since the overall structure of the course included lectures, debriefs, and demonstrations in addition to PSS, the survey included questions about what part of the class was preferable.
The options of lecture, PSS, both, or neither were given as choices.
The lecture option refers to all the class periods that were not PSS sessions.
The results can be found in Table \ref{tab:201_pref}.
Thankfully only one student selected ``neither'' across the three years.
Only 5.93\% preferred the lectures, while 39.14\% preferred just PSS, and 55.08\% preferred the combination of both.
These results suggest the importance of PSS over simply lecturing but also the need for some combination of the two.
The demonstrations and debrief lectures were important for the students to learn both essential aspects of the concepts and techniques as well as a chance to see how an expert would use them.
The combination of student application of knowledge through PSS contrasted with an expert's demonstration does match the model of apprenticeship learning better than student application alone.
The importance of the two is reflected in the survey results.

A recent pair programming meta-study concluded, ``There is little systematic knowledge from meta-analyses or literature reviews on effective instructional design for pair programming, which in fact is a base for effective learning" \citep{hawlitschek2022empirical}.
The results suggest that the combination of problem solving plus dynamic scaffolding and the weekly demonstration-PSS-debrief structure, are an effective instructional design for educational pair programming.
PSS for CS combines both the benefits of active learning along with enough guidance to maintain a cohesive, effective pair.
As the pair programming meta-study observed, ``students – at least, novices – usually need instructional support to ensure the quality and success of collaborative learning activities" \citep{hawlitschek2022empirical}.
PSS for CS supplies this supporting instructional structure.

Our results also indicate that students generally found working with a partner to be beneficial.
On average, 94.92\% of the respondents indicated that their partner was useful.
This reflects other studies which found that working in pairs was more enjoyable than working alone \citep{williams2000strengthening}, enhanced their learning \citep{williams2001support}, and lead to more persistence and success \citep{mcdowell2002effects, hannay2009effectiveness}.

The majority of research on course design for pair programming has focused on how to form the pairs.
Criteria have included student confidence, prior experience, genders, consistency of mental models, or other personality traits
However the results of these studies were inconclusive to how best form the student pairs \citep{hawlitschek2022empirical}.
Our use of random assignment for pair selection for each PSS session appears to have worked well according to the results of Table \ref{tab:201}.
Altogether this might suggest that the most effective pairing strategy is to vary the pairings.
This aligns with the notion of distributed learning, that learning is facilitated by the interactive of a variety of perspectives.



\section{Conclusion}

Pair programming has long been a promising pedagogical tool but its application to the classroom has seen mixed results.
In particular, instruction design for pair programming has seen little research.
PSS however is a natural fit for both CS education and for pair programming in particular.

PSS is an active learning pedagogy that involves student pairs solving problems in class.
It uses dynamic scaffolding to adjust the problem difficulty to match student ability in order to keep them in their ZPD.
PSS is an apprenticeship model of learning that has been successful in engineering education.

The adaptation of PSS for CS presented and studied here appears to be a good solution to the problem of applying pair programming to the classroom
Both the ``problem ladder'' and dynamic scaffolding provide enough guidance and direction for students of a variety of backgrounds and abilities. 
The active and adaptive nature of the learning environment resulted in a large number of students reporting to be engaged and challenged.

Further, students reported appreciating the weekly structure of the class as well.
By rotating through, demonstration, PSS, and debriefing, students were able to see and apply new concepts each week.
This provides the freedom and engagement of active learning while avoiding the pitfall of too little guidance for inexperienced or weak students.

PSS for CS combines a free and active learning environment with a deliberate structure to keep students on track.
This a fruitful middle ground we believe students find refreshing while being highly educational.
The highly positive results from students, with the overwhelming majority finding PSS for CS useful, engaging, and challenging, should encourage other educators to adapt it to their classroom.

\section{Acknowledgements}

We thank Joe Le Doux for teaching us PSS at a KEEN workshop and for his insights, comments, and edits when preparing this manuscript.
We would also like to thank the reviewers at ASEE for their thorough and helpful feedback.

\bibliography{ed}

\end{document}